\documentclass[twocolumn,showpacs,preprintnumbers,amsmath,amssymb]{revtex4}

\usepackage{graphicx}
\usepackage{dcolumn}
\usepackage{bm}

\begin{document}
\preprint{APS}

\title{Fragmentation Production of Triply Heavy Baryons at the LHC}

\author{M.A. Gomshi Nobary}
 \altaffiliation[Also at: ]{The Center for
Theoretical Physics and Mathematics, A.E.O.I., Chamran Building,
P.O. Box 11365-8486 Tehran, Iran; mnobary@aeoi.org.ir.
}
\author{R. Sepahvand}%
\altaffiliation[E-mail Address: ]{sepahvand@razi.ac.ir
}

\affiliation{Department of Physics, Faculty of Science, Razi
University, Kermanshah, Iran.}

\date{\today}

\begin{abstract}
 The triply heavy baryons in the standard model
formed in direct $c$ and $b$ quark fragmentation are the
$\Omega_{ccc}$, $\Omega_{ccb}$, $\Omega_{cbb}$ and $\Omega_{bbb}$
baryons. We calculate their fragmentation functions in leading
order of perturbative QCD. The universal fragmentation
probabilities fall within the range of $10^{-5}-10^{-7}$.We also
evaluate their cross section at the LHC ($\sqrt{s}=14$ TeV) using
next-to-leading order matrix elements for heavy quark-antiquark
pair production. We present the differential cross sections as
functions of the transverse momentum as well as the total cross
sections. They range from a few nb to a few pb.
\end{abstract}

\pacs{ 13.87.Fh, 13.85.Ni, 12.39.Hg}
\maketitle
\section {Introduction}
Heavy hadrons have been the focus of attention due to their
interesting properties. The study of production and decay of such
particles is interesting in two aspects. In the first place the
question is whether QCD is the right theory to predict the
properties of such objects through confirmation of the standard
model predictions with experimental data. Secondly this research
investigates the basic properties of the weak interactions at the
fundamental level. These states have in general a large number of
decay modes so that their observation and  measurement of their
properties require a large number of them to be produced. Their
cross section at $e^+ e^-$ collisions is very small therefore
their identification needs a messier environment of the hadronic
collider.

In the framework of the quark model, heavy baryons fall into three
categories. States containing one heavy flavor such as $\Lambda_c$
and $\Lambda_b$, are interesting states due to the fact that they
carry the original heavy flavor polarization [1]. Their production
has been studied in interesting models [2]. They are also being
studied experimentally [3]. The second category involves baryons
with two heavy flavor like the states $\Xi_{cc}$, $\Xi_{bb}$ and
$\Xi_{bc}$ [4]. They are treated within the approximate
quark-diquark model [5]. The model treats the production of the so
called diquark perturbatively similar to $B_c$ states [6]. Then,
it can be proved that the formation of a baryon out of a diquark
is almost the same as the fragmentation of an antiquark into a
meson [7]. In this way one obtains the fragmentation functions,
the total production probabilities and their event rates in a
desired collider. Indeed the light degree of freedom within these
states does not allow full perturbative calculation.

In the third category, we have baryons with three heavy
constituents. Since the top quark cannot take part in strong
interactions [8], there remains only the charm and bottom quarks
to form such baryons.There have been attempts to evaluate the
production of $\Omega_{ccc}$ and $\Omega_{bbb}$ in $e^+e^-$ and
hadron colliders in the quark-diquark model [9] and also using
perturbative QCD [10]. The results from $e^+e^-$ annihilation are
very small indeed [11]. However sizable rates are expected in
energetic hadron colliders [12]. Therefore the standard model
production rates of these bound states can be compared with
experimental data [13].

Consistent with the quark model of hadrons, the spectroscopy and
production mechanism of heavy meson and baryon states have been
treated satisfactorily. Specially the hadrons which contain $c$
and $b$ quarks or anti-quarks, are accounted for in the heavy
quark limit where the hadronic bound state is understood and the
perturbation theory is applied for the process of their
production. This has been successful in the treatment of $B_c$
states both in theory [6], and in experiment [14] and also in the
production of heavy diquarks in the treatment of doubly heavy
baryons [4]. In this work we shall apply this procedure to the
case of triply heavy baryons to obtain their fragmentation
properties and cross section at the LHC. Many of these states may
be observed at existing hadron colliders, specially at the
Tevatron, however some others have very low event rates. Therefore
we have chosen the LHC for the sake of integrity. Therefore in
this work we consider a framework which treats all triply heavy
baryons and obtain their fragmentation functions and estimate
their production at the LHC.

Our plan is as follows. In section {\bf II} we provide a general
discussion of the fragmentation process of S-wave triply heavy
baryons and calculation of their fragmentation functions. In
sections {\bf III} and {\bf IV} we calculate the fragmentation
functions for $c\rightarrow \Omega_{ccb}$ and $b\rightarrow
\Omega_{ccb}$ which we have chosen to be the basic ones such that
the other functions could be obtained from them by appropriate
choices of quark masses and other baryon characteristic
parameters. The inclusive production of these states at the LHC is
studied in section {\bf V}. Finally we discuss our results in
section {\bf VI}.

\section {Fragmentation of Triply heavy baryons}

The fragmentation functions are process independent and can be
applied to the $e^+ e^-$, partonic and hadronic production
processes.  At sufficiently large transverse momenta, the dominant
production mechanism is actually the fragmentation, the production
of a parton with high transverse momentum which subsequently
splits into a triply heavy state and other partons. Fig.1 shows
the fragmentation of a heavy quark $Q$ into a triply heavy baryon
$B(QQ'Q'')$ in lowest order perturbation theory. We will calculate
such Feynman diagrams.

\begin{figure}
\begin{center}
\includegraphics[width=13 cm]{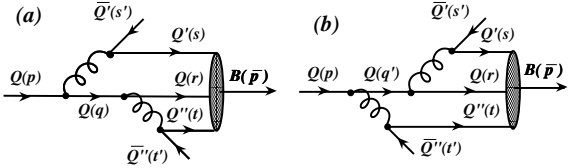}
\end{center}
\caption{ The lowest order Feynman diagrams contributing to the
fragmentation of a heavy quark ($Q$) into a triply heavy baryon
(B). The four momenta are labelled.}
\end{figure}

 The fragmentation of a
parton into a baryon state is described by fragmentation function
$D(z, \mu)$, where $z$ is the longitudinal momentum fraction of
the baryon state and $\mu$ is a fragmentation scale.  The
fragmentation function for the production of an S-wave triply
heavy baryon $B$ in the fragmentation of a quark $Q$ is obtained
from [15]

\begin{eqnarray}
D_Q^B(z,\mu_\circ)&=&\frac{1}{2}\sum_s \int \vert T_B
\vert^2\nonumber\\
 &&\times\delta^3( \overline{{\bf p}}+{\bf s'}+{\bf t'}-{\bf p})
{\rm d^3 \overline{{ p}}}{\rm d^3 { s'}}{\rm d^3 { t'}},
\end{eqnarray}
 where four momenta are as labelled in Fig 1. $T_B$ is
 the amplitude of the baryon production which
involves the hard scattering amplitude $T_H$ and the
non-perturbative smearing of the bound state. The average over
initial and the sum over final spin states are assumed.

To absorb the soft behavior of the bound state into the hard
scattering amplitude we have used the scheme introduced in [16].
The probability amplitude at large momentum transfer
 factorizes into a convolution of the hard-scattering amplitude $T_H$
and baryon distribution amplitude $\phi_B$, i.e.

\begin{eqnarray}
T_B(p,\overline p,s',t')=\int\bigl[ dx \bigr] T_H(p,\overline
p,s',t',x_i) \phi_B (x_i,w^2).
\end{eqnarray}

\noindent where $\phi_B$ is the probability amplitude to find
quarks co-linear up to a scale of $w^2$ in the baryonic bound
state. In (2),  $\bigl[ dx \bigr]=dx_1dx_2dx_3
\delta(1-x_1-x_2-x_3)$ and $x_i$'s are the momentum fractions
carried by the constituent quarks and finally $T_H$ is written in
the following form in the old fashioned perturbation theory to
keep the initial heavy quark on mass shell,

\begin{eqnarray}
T_H=\frac{16\pi^2\alpha_{s}(2m_{Q'})\alpha_{s}(2m_{Q''})MC_F }
{2\sqrt{2p_\circ \overline p_\circ s'_\circ t'_\circ}} \frac
{\Gamma} {(\overline p_\circ+s'_\circ+t'_\circ-p_\circ)}.
\end{eqnarray}

\noindent Here $\Gamma$ represents an appropriate combination of
the propagators and the spinorial parts of the amplitude.
$\alpha_s=g^2/4\pi$ is the strong interaction coupling constant.
 $C_F$ is the color factor and $M=mm_Q
m_{Q'} m_{Q''}$ with $m$ being the baryon mass. Since we ignore
the virtual motion of the baryon constituents, we propose a delta
function type function to represent the probability amplitude of
the baryon state

\begin{eqnarray}
\phi_B= f_B\; \delta\biggl\{x_i-\frac{m_i}{ m}\biggr\},
\end{eqnarray}

\noindent where $f_B$ is the baryon decay constant and is
introduced similar to meson decay constant $f_M$. Putting this
expression and (3) in (2) and carrying out the necessary
integrations, we find

\begin{eqnarray}
T_B&=&\frac{16\pi^2 \alpha_{s}(2m_{Q'})\alpha_{s}(2m_{Q''})M
C_Ff_B} {2\sqrt{2p_\circ \overline p_\circ s'_\circ
t'_\circ}}\nonumber\\
&&\times \frac{\Gamma} {(\overline
p_\circ+s'_\circ+t'_\circ-p_\circ)}.
\end{eqnarray}
With this amplitude we find the fragmentation function as
\begin{widetext}
\begin{eqnarray}
D_Q^B(z,\mu)=32\bigl[\pi^2\alpha_s({2m_{Q'}})\alpha_s({2m_{Q''}})MC_Ff_B\bigr]^2
 \int \frac{ \frac{1}{ 2}\sum\overline \Gamma \Gamma \delta^3(\overline
{\bf p}+{\bf s'}+{\bf t'}-{\bf p})} {p_\circ \overline p_\circ
s'_\circ t'_\circ(\overline p_\circ+s'_\circ+t'_\circ-p_\circ)^2}
{\rm d^3} \overline { p}{\rm d^3 }{s'}{\rm d^3 }{ t'}.
\end{eqnarray}
\end{widetext}

To proceed we need to specify our kinematics.\vskip 1 cm

We let the baryon move in the $z$ direction after production,
neglecting the virtual motion of the constituents. The initial
state heavy quark has a transverse momentum which should be
carried by the two antiquarks away through the final state jet. We
have assumed that there will be only one jet in the final state.
This assumption is justified due to the fact that the very high
momentum of the initial heavy quark will predominantly be carried
in the forward direction. Due to momentum conservation, the total
transverse momentum of the two jets will be identical to the
transverse momentum of the initial heavy quark. Therefore the
antiquark's contributions to this jet are assumed to be
proportional to their masses.

The fragmentation parameter $z$, is defined as usual, i.e.

\begin{eqnarray}
z=\frac{(E+p_\Vert)_B} {(E+p_\Vert)_Q},
\end{eqnarray}
which reduces to the following in the infinite momentum frame
which we have adopted for our study
\begin{eqnarray}
z=\frac{E_B} {E_Q}.
\end{eqnarray}

Now we set up our kinematics. According to Fig. 1 the baryon takes
a fraction $z$ of the initial heavy quark's energy (each
constituent a fraction of $x_1$, $x_2$ and $x_3$) and the two
anti-quarks take the remaining $1-z$ ($x_4$ and $x_5$ each). Thus
the four momenta of the particles are parameterized as
\begin{eqnarray}
\overline p_\circ&=&zp_\circ\;\;\;s_\circ=x_1zp_\circ\;\;\;
r_\circ=x_2zp_\circ\;\;\; t_\circ=x_3zp_\circ\nonumber\\
s'_\circ&=&x_4(1-z)p_\circ\;\;\; t'_\circ=x_5(1-z)p_\circ\;\;\; ,
\end{eqnarray}
where the condition $x_1+x_2+x_3=1$ holds. Moreover regarding our
assumptions we have
\begin{eqnarray}
{\bf s'}_T=x_4{\bf p}_T\;\;\;\ {\bf t'}_T=x_5{\bf p}_T,
\end{eqnarray}
along with the constraint of $x_4+x_5=1$.

Fig. 2 shows the lowest order Feynman diagrams for the
fragmentation of $\Omega_{ccc}$ and $\Omega_{bbb}$ (a,b),
$\Omega_{ccb}$ in the $c$ quark fragmentation (c,d) and
$\Omega_{ccb}$ in the $b$ quark fragmentation (e,f). There are
similar diagrams contributing to $\Omega_{cbb}$ fragmentation in
$b$ and $c$ quark fragmentation which are simply obtained by
interchanging the $c$ and $b$ quarks in (c,d) and (e,f)
respectively. Let us first consider the case of $c\rightarrow
\Omega_{ccb}$.

\begin{figure}
\begin{center}
\includegraphics[width=13 cm]{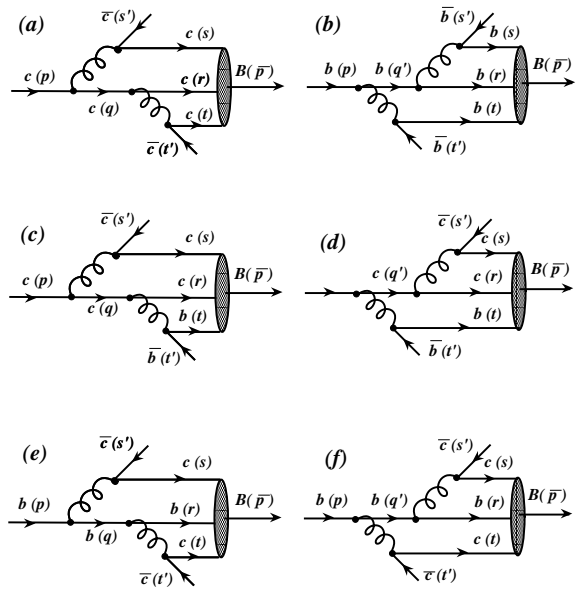}
\end{center}
\caption{ The lowest order Feynman diagrams which contribute to
different $\Omega$ baryon production. While (a) and (b) shows $c$
and $b$ quark fragmentation into $\Omega_{ccc}$ and $\Omega_{bbb}$
respectively, (c) and (d)contribute to $\Omega_{ccb}$ production
in $c$ and (e) and (f) in $b$ quark fragmentation to the same
state. Interchange of $c\longleftrightarrow b$ in last two pairs
give the contributing diagrams to $\Omega_{cbb}$ in $b$ and
$\Omega_{cbb}$ in $c$ quark fragmentation respectively.}
\end{figure}
\section{ $\Omega_{ccb}$ in $c$ quark fragmentation}
Here the diagrams (c) and (d) in Fig 2 are relevant. In each
diagram there are three propagators. In this specific case for the
diagram (c) we find the combination
\begin{eqnarray}
G&=&\frac{1}{8m_1^2m_2^3m' f^3(z)},
\end{eqnarray}
where one third of the power of $f$ comes from each propagator,
$m'=(m_1+m_2)$ and that
\begin{eqnarray}
f(z)=1+\frac{m}{ 2m'}\frac{1-z}{ z}+\frac{m'}
{2m}\Bigl(1+\frac{p_T^2}{m'^2}\Bigr)\frac{z} {(1-z)}.
\end{eqnarray}

In the case of diagram (d) we have
\begin{eqnarray}
G'=\frac{1}{16m_1^4m_2^2 f^3(z)}.
\end{eqnarray}

We put the dot products of the relevant four vectors in the
following form
\begin{eqnarray}
s'.r=m_1^2\alpha\;\;\;r.p=m_1^2\beta\;\;\;s'.p=m_1^2\gamma
\end{eqnarray}
where
\begin{eqnarray}
\alpha=\frac{1}{ 2}\biggl[\frac{m}{m'}\frac{1-z}{ z}
+\frac{m'}{m}\Bigl(1+\frac{p_T^2}{m'^2}\Bigr)\frac{z}{1-z}\biggr],
\end{eqnarray}

\begin{eqnarray}
\beta=\frac{1}{2}\biggl[\frac{m_1}{ m}\Bigl(1+\frac{p_T^2}{ m_1^2}
\Bigr)z+\frac{m}{ m_1}\frac{1}{ z}\biggr],
\end{eqnarray}
\begin{eqnarray}
\gamma&=&\frac{1}{2}\biggl[\frac{m_1}{ m'}\Bigl(1+\frac{p_T^2}
{m_1^2}
\Bigr)(1-z)\nonumber\\
&&+\frac{m'}{m_1}
\Bigl(1+\frac{p_T^2}{m'^2}\Bigr)\frac{1}{1-z}-\frac{2p_T^2}{m_1m'}\biggr].
\end{eqnarray}
In obtaining the above results we have used (9) and (10) with
$x_1=x_2=m_1/m\,;x_3=m_2/m,\;x_4=m_1/m'$ and $x_5=m_2/m'$.

In this case the $\Gamma$ in (6) reads
\begin{widetext}
\begin{eqnarray}
\Gamma=\;G\overline u(t) \gamma^\rho v(t') \Bigl[ \overline
u(r)\gamma_\rho (\rlap / q +m_1)\gamma_\nu u(p)\Bigr]\overline
u(s) \gamma^\nu v(s')+ G'\overline u(s) \gamma^\nu v(s') \Bigl[
\overline u(r)\gamma_\nu (\rlap / q' +m_1)\gamma_\rho
u(p)\Bigr]\overline u(t) \gamma^\rho v(t').
\end{eqnarray}

From which we find

\begin{eqnarray}
\overline\Gamma \Gamma&=&\;\;G^2\Bigl[(\rlap /
s'-m_1)\gamma^\mu(\rlap
/s+m_1)\gamma^\nu\Bigr]T_{\mu\sigma\rho\nu}\Bigl[(\rlap /
t'-m_2)\gamma^\sigma(\rlap / t+m_2)\gamma^\rho\Bigr]+ \nonumber\\
&&\;G'^2\Bigl[(\rlap / t'-m_2)\gamma^\sigma(\rlap
/t+m_2)\gamma^\rho\Bigr]T_{\sigma \mu\nu \rho}\Bigl[(\rlap /
s'-m_1)\gamma^\mu(\rlap / s+m_1)\gamma^\nu\Bigr]+\nonumber\\
&&GG'\Bigl[(\rlap / s'-m_1)\gamma^\mu(\rlap
/s+m_1)\gamma^\nu\Bigr]T_{\mu\sigma\nu\rho}\Bigl[(\rlap /
t'-m_2)\gamma^\sigma(\rlap / t+m_2)\gamma^\rho\Bigr]+\nonumber\\
&&GG'\Bigl[(\rlap / t'-m_2)\gamma^\sigma(\rlap
/t+m_2)\gamma^\rho\Bigr]T_{\sigma \mu\rho\nu}\Bigl[(\rlap /
s'-m_1)\gamma^\mu(\rlap / s+m_1)\gamma^\nu\Bigr],
\end{eqnarray}

where
\begin{eqnarray}
T_{\mu\sigma\rho\nu}=(\rlap / p+m_1)\gamma_\mu (\rlap /
q+m_1)\gamma_\sigma(\rlap / r+m_1)\gamma_\rho (\rlap /
q+m_1)\gamma_\nu,
\end{eqnarray}
\begin{eqnarray}
T_{\sigma\mu\nu\rho}=(\rlap / p+m_1)\gamma_\sigma (\rlap /
q'+m_1)\gamma_\mu(\rlap / r+m_1)\gamma_\nu (\rlap /
q'+m_1)\gamma_\rho,
\end{eqnarray}
\begin{eqnarray}
T_{\sigma\mu\rho\nu}=(\rlap / p+m_1)\gamma_\mu (\rlap /
q'+m_1)\gamma_\sigma(\rlap / r+m_1)\gamma_\nu (\rlap /
q+m_1)\gamma_\rho,
\end{eqnarray}
\begin{eqnarray}
T_{\sigma\mu\rho\nu}=(\rlap / p+m_1)\gamma_\sigma (\rlap /
q+m_1)\gamma_\mu(\rlap / r+m_1)\gamma_\rho (\rlap /
q'+m_1)\gamma_\nu.
\end{eqnarray}
\end{widetext}
Next we consider the phase space integrations. Note that
\begin{eqnarray}
I=\int \frac{ \delta^3(\overline{\bf p}+{\bf s'}+{\bf t'}-{\bf
p})} {p_\circ(\overline p_\circ+k_\circ+k'_\circ-p_\circ)^2} {\rm
d}^3\overline{ p}
 =\frac{ p_\circ}{\bigl[m m'g(z)\bigr]^2},
\end{eqnarray}

\noindent where

\begin{eqnarray}
g(z)&=& -\frac{p_T^2}{m m'}+\frac{m}{m'}\frac{1}
{z}\nonumber \\
&&+\frac{m'}{m}\Bigl(1+\frac{p_T^2} {m'^2}\Bigr)\frac{1}{1-z}.
\end{eqnarray}
Here instead of performing transverse momentum integrations we
replace the integration variable by its average value in each
case. Therefore we write

\begin{eqnarray}
\int F(z,{\bf s'}_T)d^3 s'&=&\int F(z,{\bf s'}_T){\rm d}s'_L {\rm
d}^2 s_T\nonumber\\
&& =m_1^2s'_\circ F(z,\langle {s'_T}^2\rangle )
\nonumber\\
&&=m_1^2s'_\circ F\Bigl[z,\frac{m_1}{m'}\langle {p_T}^2\rangle
\Bigr],
\end{eqnarray}
and
\begin{eqnarray}
\int F'(z,{\bf t'}_T)d^3 t'&=&\int F'(z,{\bf t'}_T){\rm d}t'_L
{\rm
d}^2 t'_T \nonumber\\
&&=m_2^2t'_\circ F'(z,\langle {t'_T}^2\rangle )
\nonumber\\
&&=m_2^2t'_\circ F'\Bigl[z,\frac{m_2}{m'}\langle {p_T}^2\rangle
\Bigr].
\end{eqnarray}

Putting all this together back in (6), we obtain the fragmentation
function for $c\rightarrow \Omega_{ccb}$ as follows
\begin{widetext}
\begin{eqnarray}
D_{c\rightarrow
\Omega_{ccb}}(z,\mu_\circ)&=&\frac{\pi^4\alpha_{s}(2m_1)^2\alpha_{s}(2m_2)^2f_B^
2 C_F^2m_2^2}{128 a^2 m'^4 zf^6(z)g^2(z)}\biggl\{
6-\beta-\gamma+\alpha^3(1+\gamma)
+4a^4(\alpha-1)\Bigl[-2+\alpha(\beta-1)+\beta+\gamma\Bigr]\nonumber\
\\
&&+\alpha^2(4+5\beta+3\gamma)+\alpha(7+6\beta+5\gamma)
+8a^3\Bigl[5+\beta+\alpha^2(6+9\beta)+\alpha(7+7\beta+\gamma)\Bigr]\
\nonumber\\
&&+2a\Bigl[66+23\beta+27\gamma+\alpha^3(1+\gamma)
+\alpha^2(52+37\beta+19\gamma)+\alpha(115+70\beta+57\gamma)\Bigr]\nonumber\\
&&+a^2\Bigl[182+71\beta+71\gamma+5\alpha^3(1+\gamma)
+\alpha^2(168+141\beta+47\gamma)
+\alpha(311+198\beta+133\gamma)\Bigr]\biggr\}.
\end{eqnarray}
\end{widetext} Here $f(z)$ given by (12) is due to the propagators
and $g(z) $comes from the energy denominator (25). $\alpha$,
$\beta$ and $\gamma$ are for dot products given by (15)-(17). We
have set $a=m_1/m_2$.

It is clear that the interchange of $c \leftrightarrow b$ in the
above function will provide the fragmentation function for $ b
\rightarrow \Omega_{cbb}$ in agreement with our direct
calculation.

\section{$\Omega_{ccb}$ in $b$ quark fragmentation}

Now let us consider the process of $b \rightarrow \Omega_{ccb}$.
In this case regarding the diagrams (e) and (f) in Fig. 2 and
using the above procedure we find for the propagators
\begin{eqnarray}
G=G'=\frac{1} {8m_1^5m'f'^3(z)},
\end{eqnarray}
where $f'(z)$ is
\begin{eqnarray}
f'(z)=1+\frac{m}{4m_1}\frac{1-z}{ z}+\frac{m_1}{
m}\Bigl(1+\frac{1}{ 4} \frac{p_T^2}{m_1^2}\Bigr)\frac{z}{1-z}.
\end{eqnarray}

The dot products of the relevant four vectors are put in the
following form
\begin{eqnarray}
s'.r=m_2^2\alpha'\;\;\;r.p=m_2^2\beta'\;\;\;s'.p=m_1^2\gamma'
\end{eqnarray}
where

\begin{eqnarray}
\alpha'=\frac{1}{
2}\biggl[\frac{m}{2m_2}\frac{1-z}{z}+\frac{m_1^2} {m m_2}
\Bigl(1+\frac{1}{4}\frac{p_T^2}
{m_1^2}\Bigr)\frac{2z}{1-z}\biggr],
\end{eqnarray}

\begin{eqnarray}
\beta'=\frac{1}{ 2}\biggl[\frac{m_2}{
m}\Bigl(1+\frac{p_T^2}{m_2^2} \Bigr)z+\frac{m}{m_2}\frac{1}
{z}\biggr],
\end{eqnarray}

\begin{eqnarray}
\gamma'&=&\frac{1}{ 2}\biggl[\frac{1}{2}\Bigl(\frac{m_2}
{m_1}\Bigr)^2\Bigl(1+\frac{p_T^2}{ m_2^2
}\Bigr)(1-z)\nonumber\\
&&+\Bigl(1+\frac{1}{ 4} \frac{p_T^2}
{m_1^2}\Bigr)\frac{2}{1-z}-\frac{p_T^2}{ m_1^2}\biggr].
\end{eqnarray}
Note that the $x$'s in (9) read as $x_1=x_3=m_1/m,\;x_2=m_2/m$ and
$x_4=x_5=1/2$ in this case. Here the $\Gamma$ in (6) has the
following form
\begin{widetext}
\begin{eqnarray}
\Gamma '=\;G\overline u(t) \gamma^\rho v(t')\Bigl[ \overline
u(r)\gamma_\rho (\rlap / q +m_2)\gamma_\nu u(p)\Bigr]\overline
u(s) \gamma^\nu v(s')+G'\overline u(s) \gamma^\nu v(s') \Bigl[
\overline u(r)\gamma_\nu (\rlap / q' +m_2)\gamma_\rho
u(p)\Bigr]\overline u(t) \gamma^\rho v(t').
\end{eqnarray}
Therefore

\begin{eqnarray}
\overline\Gamma' \Gamma'&=&\;\;G^2\Bigl[(\rlap /
s'-m_1)\gamma^\mu(\rlap
/s+m_1)\gamma^\nu\Bigr]T_{\mu\sigma\rho\nu}\Bigl[(\rlap /
t'-m_1)\gamma^\sigma(\rlap / t+m_1)\gamma^\rho\Bigr]+ \nonumber\\
&&\;G'^2\Bigl[(\rlap / t'-m_1)\gamma^\sigma(\rlap
/t+m_1)\gamma^\rho\Bigr]T_{\sigma \mu\nu \rho}\Bigl[(\rlap /
s'-m_1)\gamma^\mu(\rlap / s+m_1)\gamma^\nu\Bigr]+\nonumber\\
&&GG'\Bigl[(\rlap / s'-m_1)\gamma^\mu(\rlap
/s+m_1)\gamma^\nu\Bigr]T_{\mu\sigma\nu\rho}\Bigl[(\rlap /
t'-m_1)\gamma^\sigma(\rlap / t+m_1)\gamma^\rho\Bigr]+\nonumber\\
&&GG'\Bigl[(\rlap / t'-m_1)\gamma^\sigma(\rlap
/t+m_1)\gamma^\rho\Bigr]T_{\sigma \mu\rho\nu}\Bigl[(\rlap /
s'-m_1)\gamma^\mu(\rlap / s+m_1)\gamma^\nu\Bigr],
\end{eqnarray}
where
\begin{eqnarray}
T'_{\mu\sigma\rho\nu}=(\rlap / p+m_2)\gamma_\mu (\rlap /
q+m_2)\gamma_\sigma(\rlap / r+m_2)\gamma_\rho (\rlap /
q+m_2)\gamma_\nu,
\end{eqnarray}
\begin{eqnarray}
T'_{\sigma\mu\nu\rho}=(\rlap / p+m_2)\gamma_\sigma (\rlap /
q'+m_2)\gamma_\mu(\rlap / r+m_2)\gamma_\nu (\rlap /
q'+m_2)\gamma_\rho,
\end{eqnarray}
\begin{eqnarray}
T'_{\sigma\mu\rho\nu}=(\rlap / p+m_2)\gamma_\mu (\rlap /
q'+m_2)\gamma_\sigma(\rlap / r+m_2)\gamma_\nu (\rlap /
q+m_2)\gamma_\rho,
\end{eqnarray}
\begin{eqnarray}
T'_{\sigma\mu\rho\nu}=(\rlap / p+m_2)\gamma_\sigma (\rlap /
q+m_2)\gamma_\mu(\rlap / r+m_2)\gamma_\rho (\rlap /
q'+m_2)\gamma_\nu.
\end{eqnarray}

Finally similar to our previous treatment of $c\rightarrow
\Omega_{ccb}$, we obtain the fragmentation function for
$b\rightarrow \Omega_{ccb}$ as

\begin{eqnarray}
D_{b\rightarrow
\Omega_{ccb}}(z,\mu_\circ)&=&\frac{\pi^4\alpha_{s}(2m_1)^4
 f_B^2 C_F^2}{64a^6 m'^2
zf'^6(z)g'^2(z)} \Bigl\{3\alpha'^2(1+3\beta')
+a\alpha'\Bigl[-5+\alpha'^2-4\beta'+2\alpha'(8+9\beta')\Bigr]\nonumber\\
&&+14a^5\gamma'+2a^4(12+5\beta'+\gamma'+12\alpha'\gamma')
+a^2\Bigl[2-\beta'+\alpha'^2(21+10\beta')+\alpha'(36+34\beta'-3\gamma')\nonumber\\
&&+\alpha'^3\gamma'\Bigr]+a^3\Bigl[20+6\beta'-\gamma'+11\alpha'^2\gamma'
+4\alpha'(11+5\beta'+3\gamma')\Bigr]\Bigr\}.
\end{eqnarray}
\end{widetext} Here $g'(z)$ comes from the energy denominator
which in this case reads as
\begin{eqnarray}
g'(z)&=&-\frac{ p_T^2}{2m m_1}+\frac{m}{2
m_1}\frac{1}{ z}\nonumber\\
&&+\frac{2 m_1}{m}\Bigl(1+\frac{p_T^2}{4
m_1^2}\Bigr)\frac{1}{1-z}.
\end{eqnarray}

Again in this case the interchange of $c \longleftrightarrow b$
will provide the fragmentation function for $c\rightarrow
\Omega_{cbb}$. This also agrees with our direct calculation.

In the equal mass case where $\alpha'=\alpha$, $\beta'=\beta$,
$\gamma'=\gamma$, $f'(z)=f(z)$ and $g'(z)=g(z)$  the fragmentation
function takes the form
\begin{widetext}
\begin{eqnarray}
D_{Q\rightarrow
\Omega_{QQQ}}(z,\mu_\circ)&=&\frac{\pi^4\alpha_s(2m_Q)^4
f_B^2C_F^2 }{256 m_Q^2
zf^6(z)g^2(z)}\nonumber\\
&&\times\Bigl\{46+15\beta+15\gamma+\alpha^3(1+\gamma)
+\alpha^2(40+37\beta+11\gamma+\alpha(75+50\beta+33\gamma)\Bigr\},
\end{eqnarray}
\end{widetext}
where $Q$ may be assumed to be a $c$ or $b$ quark with $m_Q$ being
respective quark mass.

The input for the fragmentation functions (28), (41) and (43) are
quark masses, baryon decay constants and the color factor. We have
set $m_1=m_c$=1.25 GeV and $m_2=m_b$=4.25 GeV. For the decay
constant and the color factor we have taken $f_B$=0.25 GeV and
$C_F=7/6$ for all cases. The later being calculated using color
line counting rule. We have also taken $\langle p_T^2\rangle$=1
GeV which is an optimum value for this quantity.

\section {Inclusive Production Cross Section }

Theoretical calculations of the production cross section in high
energy hadron collisions are based on the idea of factorization.
Essentially this idea incorporates the short distance high energy
parton production and the long distance fragmentation process.
Here it is assumed that at high transverse momentum the inclusive
production of triply heavy baryons is factorized into convolution
of parton distribution functions, bare cross section of the
initiating heavy quark and the fragmentation function, i.e.
\begin{widetext}
\begin{eqnarray}
\frac {d\sigma}{dp_T}(p p \rightarrow \Omega_{QQ'Q''}(p_T)+
X)&=&\sum_{i,j}\int dx_1 dx_2 dz f_{i/p}(x_1,\mu)f_{j/
p}(x_2,\mu)\nonumber\\ &&\times\Bigl[ \hat\sigma(ij\rightarrow
Q(p_T/z)+X,\mu) D_{ Q\rightarrow \Omega_{QQ'Q''}}(z,\mu)\Bigr].
\end{eqnarray}
\end{widetext}
Where $f_{i,j}$ are parton distribution functions with momentum
fractions of $x_1$ and $x_2$ and $\hat\sigma$ is the heavy quark
production cross section and $D(z,\mu)$ represents the
fragmentation of the produced heavy quark into a triply heavy
baryon. Note that the equation is written in the factorization
scale $\mu$. Furthermore here the factorization, renormalization
and fragmentation scales are set to be equal. In other words we
have set the scale $\mu$ in the parton distribution functions,
subprocess cross sections and the fragmentation functions to be
the same. Meanwhile the running scale used in fragmentation
functions is set to be maximum ($\mu,\;\mu_\circ$), where
$\mu_\circ$ is the prescribed initial scale for the fragmentation
functions. We have employed the parameterization due to
Martin-Roberts-Stiriling (MRS) [17] for parton distribution
functions and have included the heavy quark production cross
section up to the order of $\alpha_s ^3$ [18]. The examples of
Feynman diagrams for the $ Q \overline Q$ pair production to the
order of $\alpha_s ^2$ and $\alpha_s ^3$ are shown in figure 3 and
4 respectively.
\begin{figure}
\begin{center}
\includegraphics[width=9 cm]{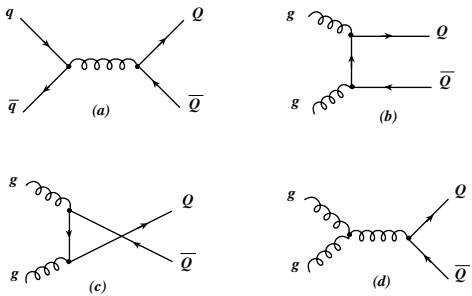}
\end{center}
\caption[Submanagers]{ Born diagrams contributing to the
calculation of heavy quark pair production cross section.}
\end{figure}
\begin{figure}
\begin{center}
\includegraphics[width=8 cm]{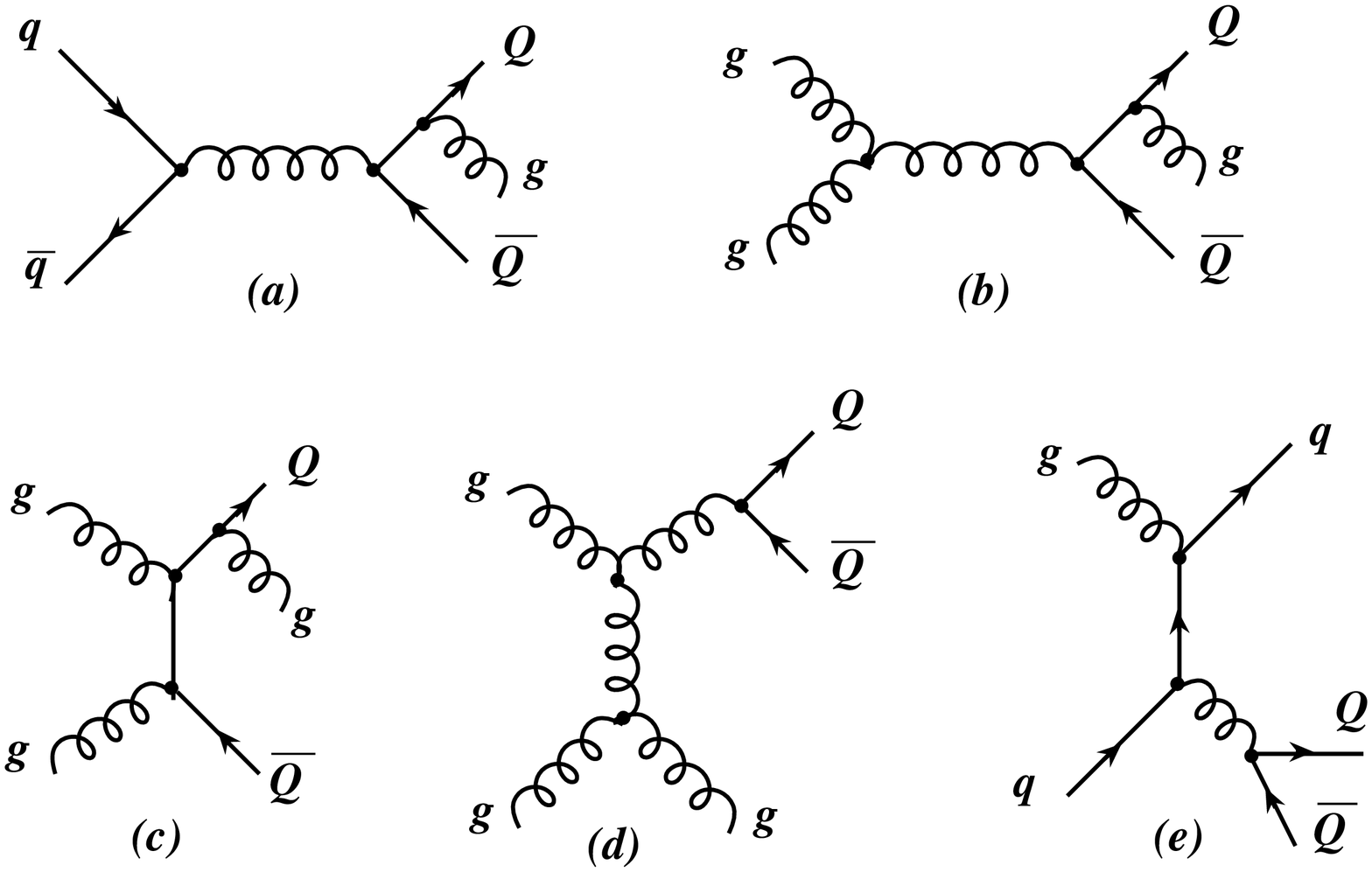}
\end{center}
\caption[Submanagers]{ Examples of Feynman diagrams up to order of
$\alpha_s^3$ contributing to quark-antiquark pair production.}
\end{figure}

 For the LHC the acceptance cuts of ${p_T}^{\rm cut} \geq 10$
GeV and  $|y|\leq 1$ are chosen where the rapidity $y$ is defined
as
\begin{eqnarray}
y=\frac{1}{2} \log\biggl\{\frac{E-p_L}{E+p_L}\biggr\}.
\end{eqnarray}

The physical production rates calculated at all orders in
perturbation theory would be independent of normalization/
factorization scale. Such results are not available. So that the
production cross sections do depend to a certain degree on the
choices of $\mu$. We will estimate the dependence on $\mu$ by
choosing the transverse mass of the heavy quark as our central
choice of scale defined by
\begin{eqnarray}
\mu_R=\sqrt{ {p_T}^2{\rm (parton)}+{m_Q}^2,   }
\end{eqnarray}
and vary it appropriately to the fragmentation scale of our
particles. This choice of scale, which is of the order of $p_T$
(parton), avoids the large logarithms in the process of the form
$\ln(m_Q/\mu)$ or $\ln(p_T/\mu)$. However, we have to sum up the
logarithms of order of $\mu_R/m_Q$ in the fragmentation functions.
But this can be implemented by evolving the fragmentation
functions by the Altarelli-Parisi equation. This equation reads as

\begin{eqnarray}
\mu \frac{\partial}{\partial \mu} D_{Q \rightarrow H}
(z,\mu)&=&\nonumber
\end{eqnarray}
\begin{eqnarray}
 \int_z^1 \frac{dy}{y} P_{Q \rightarrow Q}
(z/y,\mu) D_{Q \rightarrow H}(y,\mu).
\end{eqnarray}
\noindent Here the functions $D(z,\mu)$ at the initial scale
$\mu_\circ$ are given by (28), (41) and (43). $P_{Q \rightarrow
Q}(x=z/y,\mu) $ is the Altarelli-Parisi splitting function and at
the leading order in $\alpha_s$ reads

\begin{eqnarray}
P_{Q\rightarrow Q} (x,\mu)=\frac{4 \alpha_s(\mu)}{3 \pi}
\biggl(\frac{1+x^2}{1-x}\biggr)_+,
\end{eqnarray}
\noindent where the running coupling constant $\alpha_s(\mu)$ is
evaluated at one loop by evolving from the experimental value
$\alpha_s(M_Z)=0.1172$ [19] given by
\begin{eqnarray}
\alpha_s(\mu)=\frac{\alpha_s(M_Z)}{1+8\pi b_{\circ} \alpha_s(M_Z)
\ln (\frac{\mu }{ M_Z})}, \;\;\; b_\circ = \frac{33-2 n_f} {48
\pi^2}.
\end{eqnarray}
Here $n_f$ is number of flavors below the scale $\mu$. The (+)
prescription reads $ f(x)_+ =f(x)-\delta (1-x) \int_0^1 f(x')dx'$.
We note that only $P_{Q \rightarrow Q}$ splitting function appears
in (47). This is because the quark $ Q$ is assumed to be heavy
enough to make other contributions namely
 $P_{Q \rightarrow g}$,
 $P_{g \rightarrow Q}$ and
 $P_{g \rightarrow g}$ irrelevant.
The boundary condition on the evolution equation (47) is the
initial fragmentation function $D_{Q \rightarrow H}(z,\mu)$ at
some scale $\mu=\mu_\circ$ where its calculation is possible.
\begin{figure*}
\begin{center}
\includegraphics[width=8.5 cm]{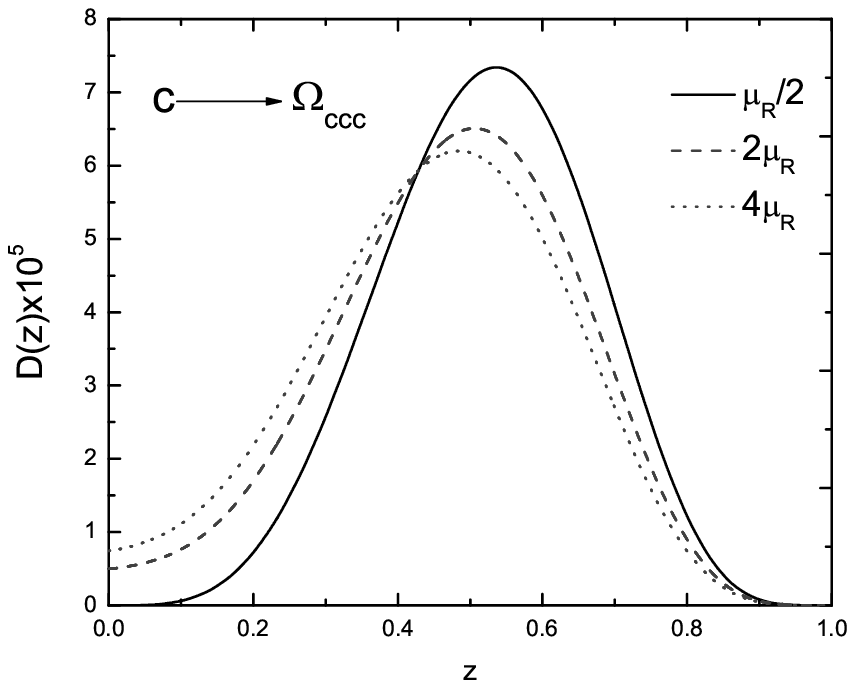}
\includegraphics[width=8.5 cm]{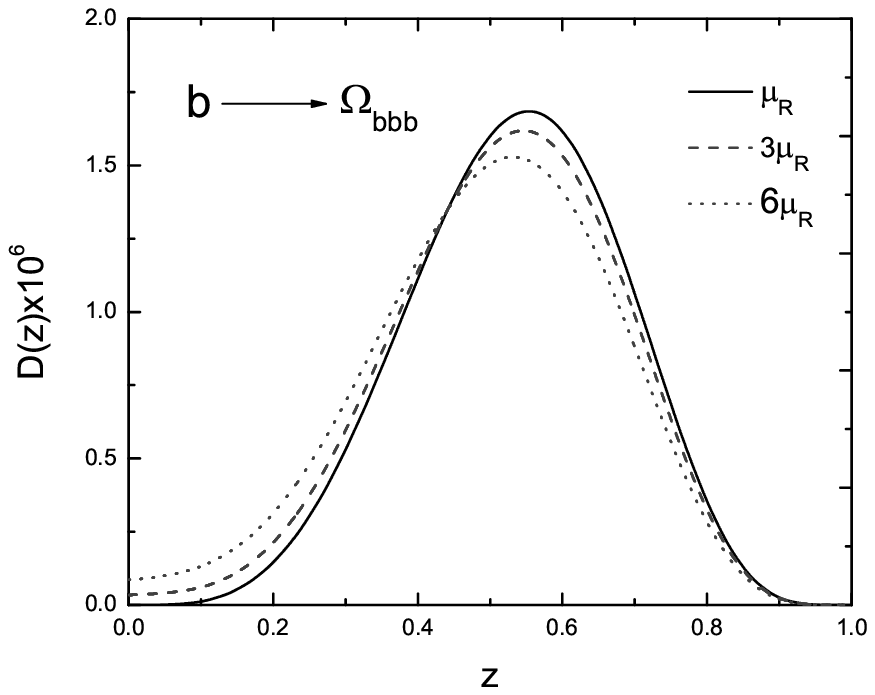}
\vskip .0cm
\includegraphics[width=8.5 cm]{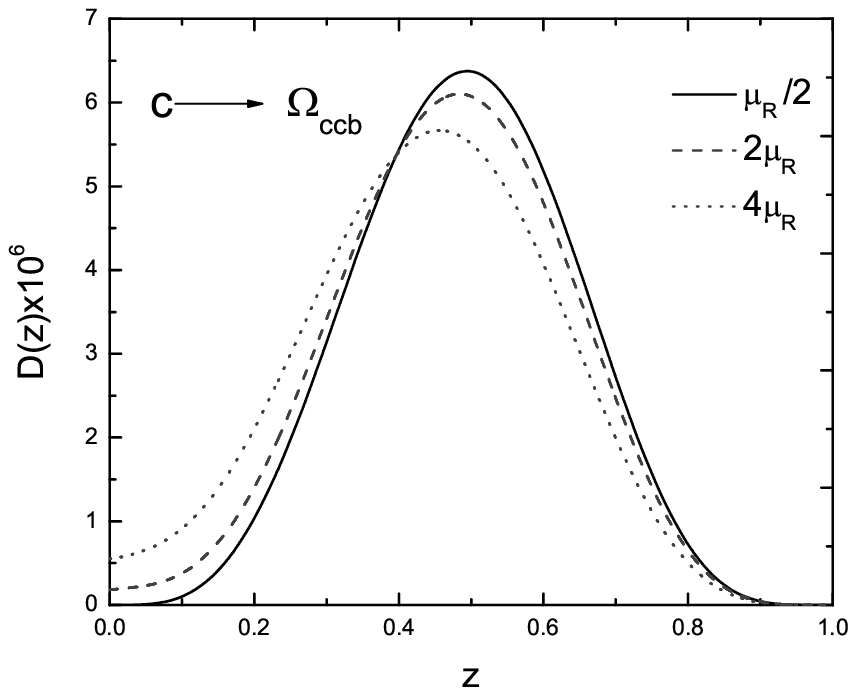}
\includegraphics[width=8.5 cm]{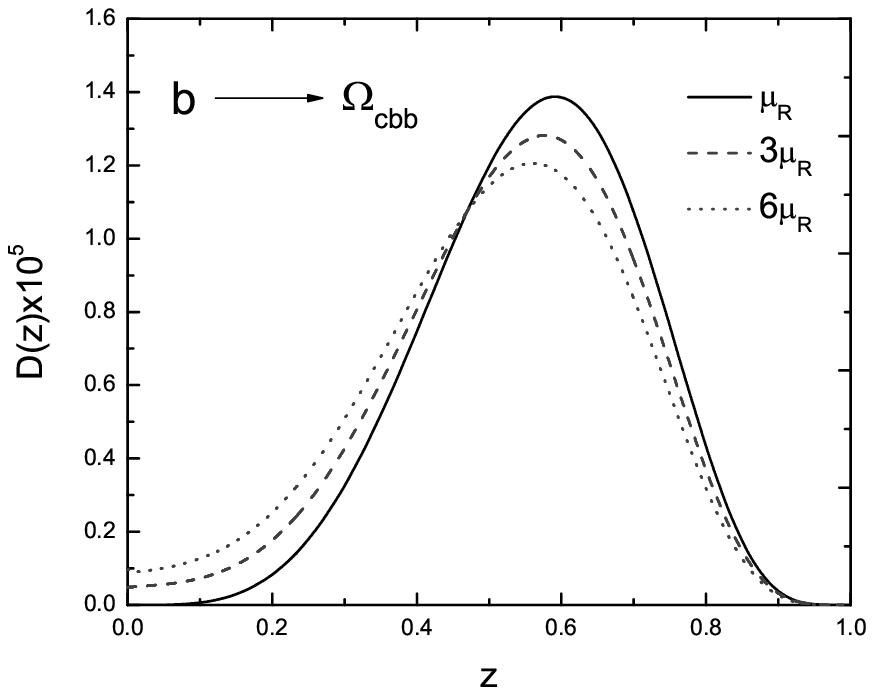}
\vskip .0cm
\includegraphics[width=8.5 cm]{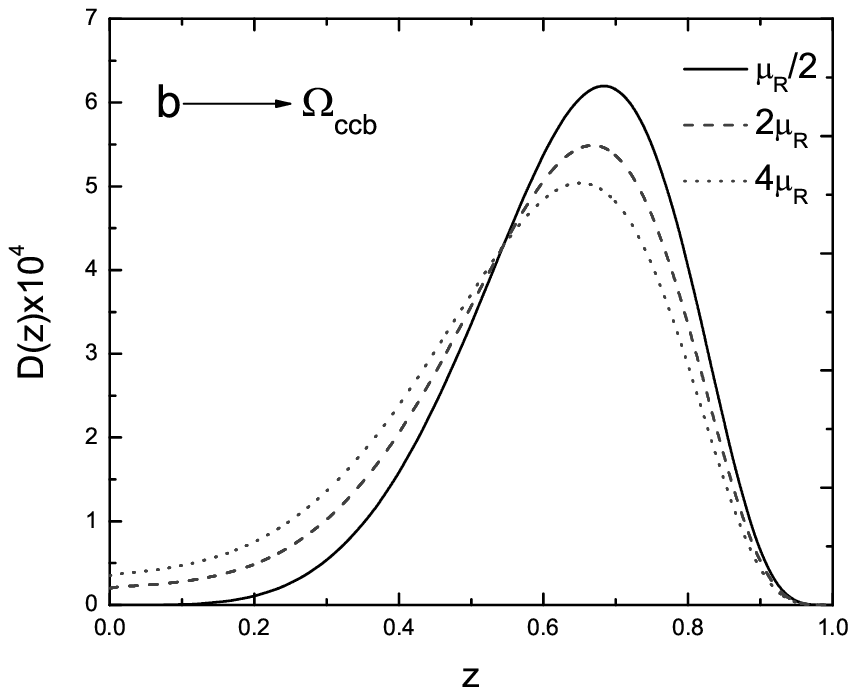}
\includegraphics[width=8.5 cm]{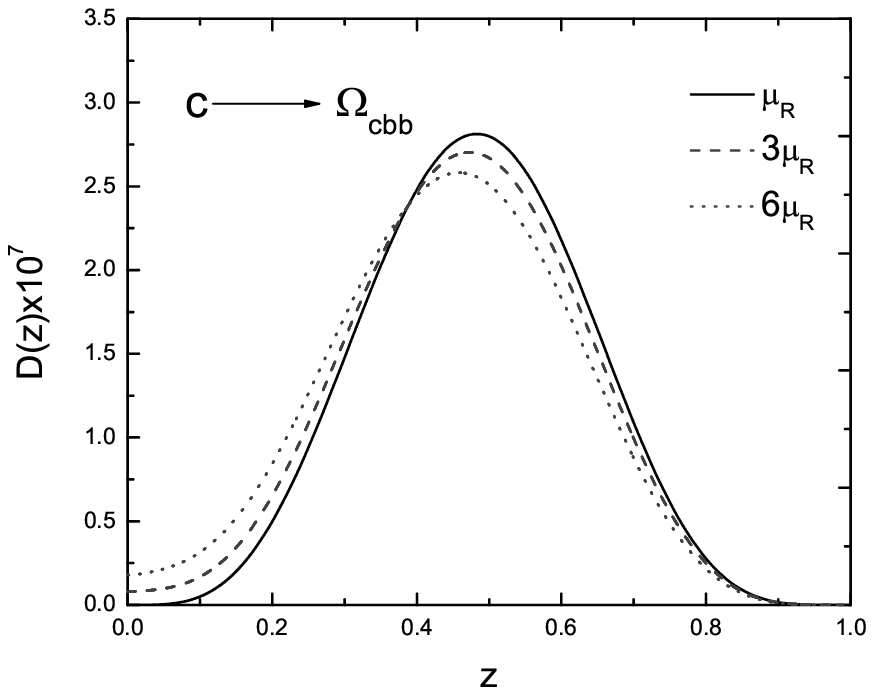}
\end{center}
\caption[Submanagers]{ Fragmentation of a $c$ and $b$ quark into
possible triply heavy baryons. Note that they are grouped
according to their heavy contents. Evolution to desired scales are
shown for the LHC. We have used two sets of scales $\mu=
\mu_R/2,\;2\mu_R,\;4\mu_R$ (left) and  $\mu=
\mu_R,\;3\mu_R,\;6\mu_R$ (right).}
\end{figure*}
\begin{figure*}
\begin{center}
\includegraphics[width=8.5 cm]{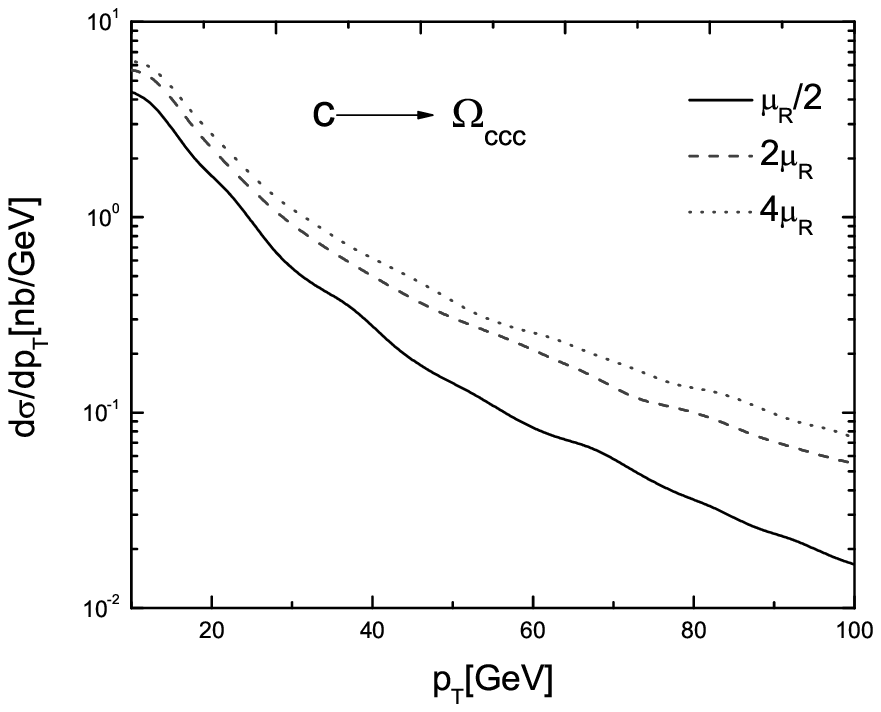}
\includegraphics[width=8.5 cm]{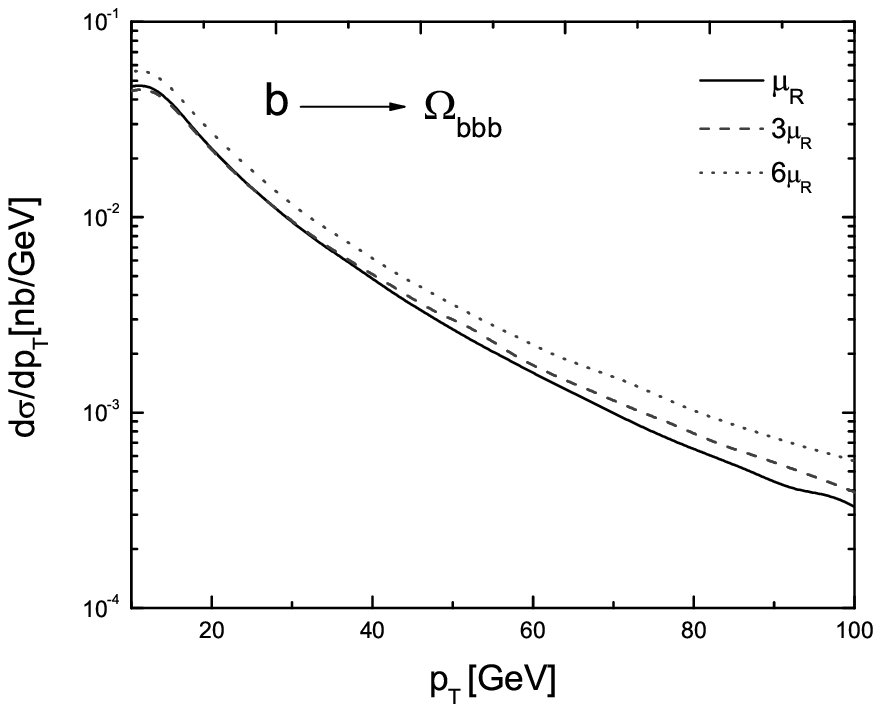}
\includegraphics[width=8.5  cm]{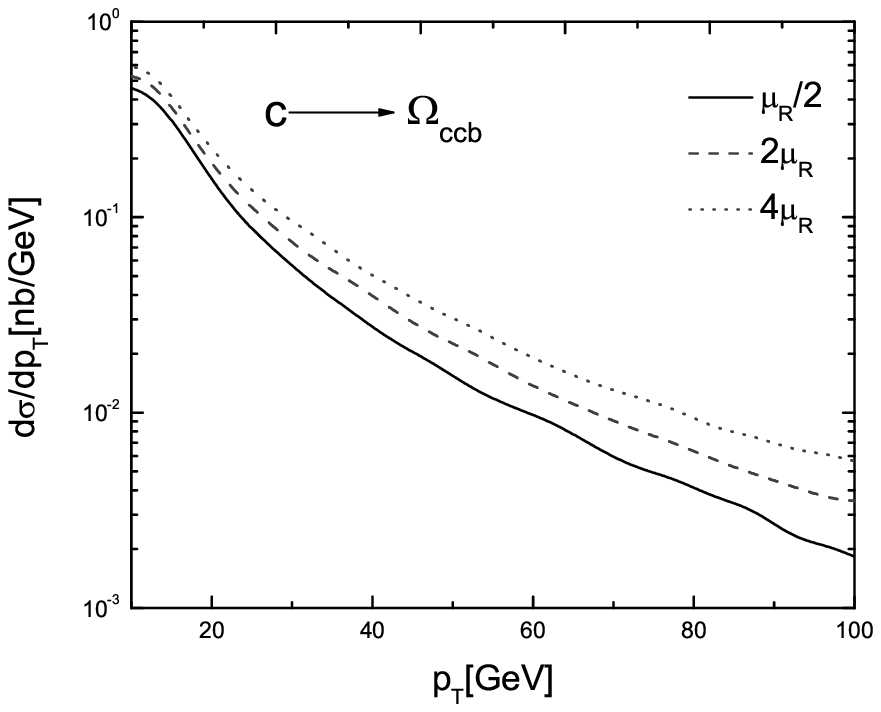}
\includegraphics[width=8.5  cm]{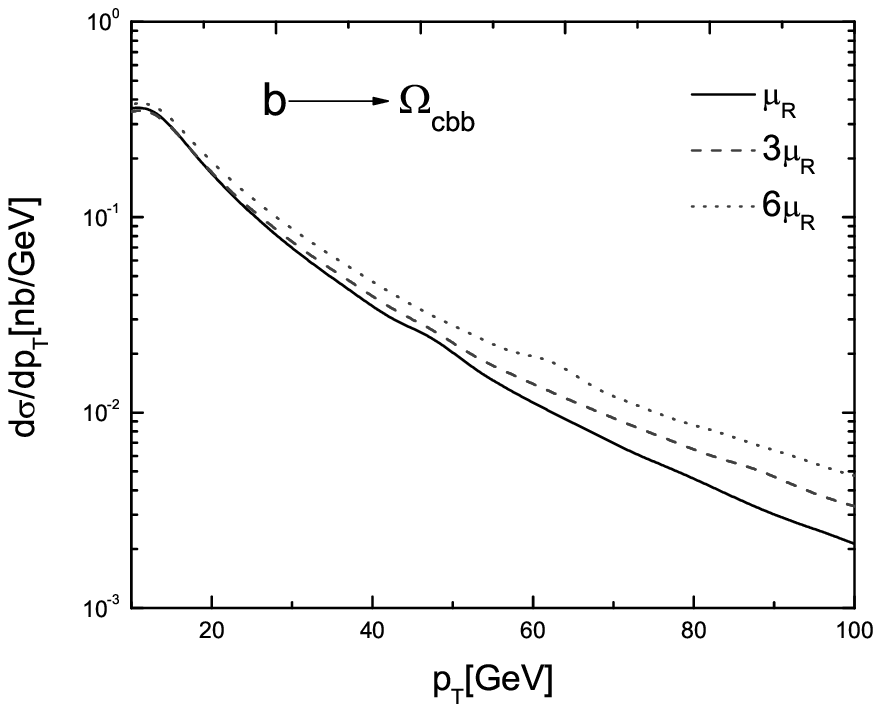}
\includegraphics[width=8.5  cm]{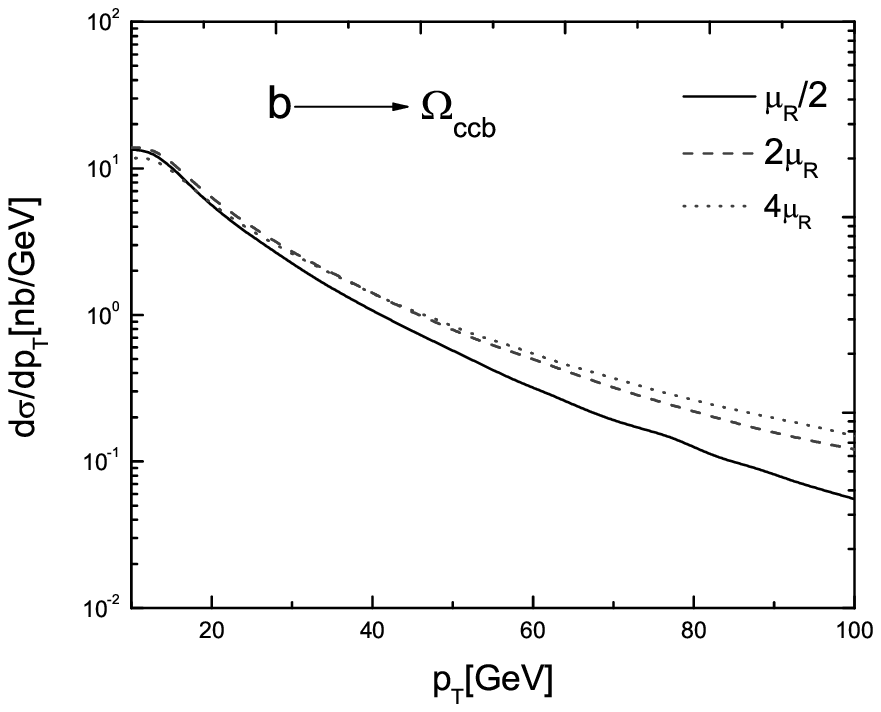}
\includegraphics[width=8.5 cm]{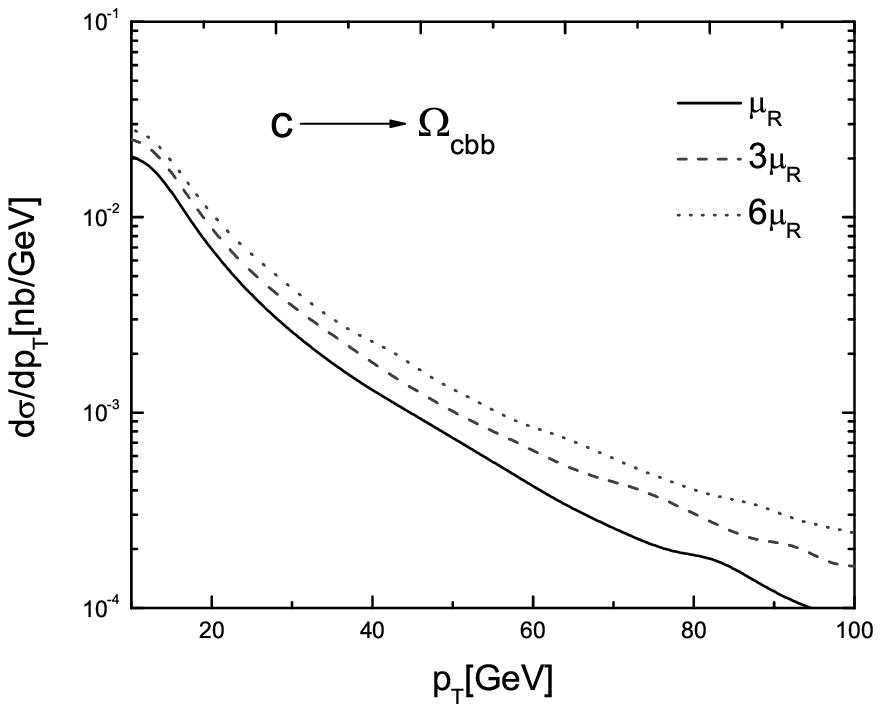}
\end{center}
 \caption[Submanagers]{ Differential cross section
${\rm d}\sigma/{\rm d}p_T[{\rm nb/(GeV)}]$ versus transverse
momentum $p_T$ for production of different triply heavy baryons in
$c$ and $b$ fragmentation at the LHC. In each graph the
distribution is shown for scales specified. Note that the sets of
scale used in each column of diagrams is the same (the baryons in
the left column contain at least two $c$ and in the right column
at least two $b$ quark). The kinematical cuts imposed are
${p_T}^{\rm cut}>10$ GeV and$|y|\leq 1$. }
\end{figure*}

\section{Results and discussion}
In the heavy quark limit we have obtained exact analytical
fragmentation functions for S-wave triply heavy baryons using
leading order perturbative QCD. The non-petubative part of the
bound state is treated by employing a delta function type
distribution function thus ignoring the respective motion of the
constituents. We have obtained the fragmentation functions for $c
\rightarrow \Omega_{ccb}$ and $b \rightarrow \Omega_{ccb}$ at the
scale of $\mu_\circ=m+m_{Q'}+m_{Q''}$ where $m$ is the baryon mass
and ${Q'}$ and ${Q''}$ are specified in Fig. 1. The functions for
$c \rightarrow \Omega_{cbb}$ and $b \rightarrow \Omega_{cbb}$ are
obtained simply by the interchange of $c \leftrightarrow b$ in
agreement with direct calculations. The fragmentation functions
for $c \rightarrow \Omega_{ccc}$ and $b \rightarrow \Omega_{bbb}$
are obtained by setting the $c$ and $b$ quark masses to be equal.

With our choice of quark masses, i.e. $m_c$=1.25 GeV, $m_b$=4.25
GeV and $\mu_R$ defined by (46) the behavior of fragmentation
functions as well as the transverse momentum distributions of the
differential cross sections are well analyzed if we put the triply
heavy baryons in two groups. Those which contain at least two $c$
and those at least two $b$ quarks. Therefore while we need to
study the  $c \rightarrow \Omega_{ccc}$, $c \rightarrow
\Omega_{ccb}$, $b \rightarrow \Omega_{ccb}$ within a lower set of
$\mu= \mu_R/2,\;2\mu_R$ and $4\mu_R$ scales, $b \rightarrow
\Omega_{bbb}$, $b \rightarrow \Omega_{cbb}$, $c \rightarrow
\Omega_{cbb}$ would require higher set of $\mu= \mu_R,\;3\mu_R$
and $6\mu_R$. This also provides a means by which the sensitivity
of the results are tested. When the first (out of the three
selected) scale $\mu=\mu_R/2$ ($\mu=\mu_R$) is less than
$\mu_\circ$, which incidently happens for all of our particles, we
choose the larger of ($\mu,\;\mu_\circ$).

The behavior of our fragmentation functions along with their
evolutions at $\mu= \mu_R/2$ ($\mu_R$), $2\mu_R$ ($3\mu$ and
$4\mu_R$ ($6\mu)$ using the Altarelli-Parisi evolution equation
(47) are shown in figure 5 for different $\Omega$ states. Note
that the scales used here in each diagram are the same as the ones
which are employed in $p_T$ distribution diagrams in Fig. 6. Also
note that each column of diagrams are sketched in separate set of
scales. The universal fragmentation probabilities and the average
fragmentation parameters at $\mu_\circ$ are shown in table I. The
probabilities at this table indicate that while some of the states
would have considerable event rates at existing colliders, others
are less probable. Therefore here we present their cross sections
at the LHC at $\sqrt{s}$=14 TeV.

The differential cross sections are shown in figure 6. The slow
fall off of the distributions is expected in the framework of our
study. We are dealing with a collider with large $\sqrt {s}$ and
production of heaviest hadrons in the standard model both against
a sharp fall off. A look at this figure reveals that firstly the
differential cross sections are sensitive for different scales
chosen only at high transverse momentum for nearly all states.
secondly as the number of the $b$ quark is increased in the state,
the differential cross section is less sensitive for higher
scales. It is seen in the cases of $b\rightarrow\Omega_{bbb}$,
$b\rightarrow\Omega_{cbb}$ and $c\rightarrow\Omega_{cbb}$. This
was the main reason to choose two different sets of scales here.
Sometimes it happens that the distribution for two different
scales cross each other. This occurs for
$b\rightarrow\Omega_{bbb}$ at $p_T$=20 GeV and for
$b\rightarrow\Omega_{cbb}$ at nearly $p_T$=15 GeV and
$b\rightarrow\Omega_{ccb}$ at $p_T$=20 GeV all for the first two
scales. This means that rate of decrease in differential cross
sections are different in the specified $p_T$ region for the two
scales.
\begin{table}
\caption{The universal fragmentation probabilities (F.P.) and the
average fragmentation parameter $\langle z \rangle$ at
fragmentation scale for different $\Omega$ states in possible $c$
and $b$ quark fragmentation.}
\begin{ruledtabular}
\begin{tabular}{ccc}{\rm Process} & F.P.&
$\langle z \rangle(\mu_\circ)$\\\hline
$c\rightarrow\Omega_{ccc}$&$2.789\times 10^{-5}$&0.521\\
$c\rightarrow\Omega_{ccb}$&$2.475\times10^{-6}$&0.490\\
$b\rightarrow\Omega_{ccb}$&$2.183\times10^{-4}$&0.634\\
$b\rightarrow\Omega_{bbb}$&$6.459\times10^{-7}$&0.534\\
$b\rightarrow\Omega_{cbb}$&$5.290\times
10^{-6}$&0.562\\
$c\rightarrow\Omega_{cbb}$&$1.086\times10^{-7}$&0.482\\
\end{tabular}
\end{ruledtabular}
\end{table}

\begin{table*}
\caption{Total cross section in pb for triply heavy baryons in
possible $c$ and $b$ quark fragmentation for various scales at the
LHC with $\sqrt{s}=14 {\rm TeV}$ where the kinematical cuts of
$p_T>10$ GeV and$|y|\leq 1$ are imposed. Note that the cross
sections are calculated in two groups of scales, ($\mu=
\mu_R/2,\;2\mu_R$ and $4\mu_R$) for lighter
$c\rightarrow\Omega_{ccc}$, $c\rightarrow\Omega_{ccb}$ and
$b\rightarrow\Omega_{ccb}$ and ($\mu= \mu_R,\;3\mu_R$ and
$6\mu_R$) for heavier $c\rightarrow\Omega_{cbb}$,
$b\rightarrow\Omega_{cbb}$ and $b\rightarrow\Omega_{bbb}$ states.
The ratios $\sigma(Q\rightarrow\Omega)/\sigma(Q)$ are given in the
last column. They are calculated at $\mu=2\mu_R$. }
\begin{ruledtabular}
\begin{tabular}{cccccccc}Process of& & & & Cross Section [pb] & & &
The Ratio\\\cline {2-7}Production &$\mu_R/2$
&$\mu_R$&$2\mu_R$&$3\mu_R$
&$4\mu_R$&$6\mu_R$&$\sigma(Q\rightarrow\Omega)/\sigma(Q)$\\\hline
$c\rightarrow\Omega_{ccc}$&301.88&&306.99&&307.59&&$1.20\times 10^{-7}$\\
$c\rightarrow\Omega_{ccb}$&26.58&&30.03&&29.88&&$1.18\times10^{-8}$\\
$b\rightarrow\Omega_{ccb}$&2153.08&&2155.31&&1723.80&&$3.61\times10^{-6}$\\
$b\rightarrow\Omega_{bbb}$&&6.34&6.38&5.77&&8.40&$1.36\times10^{-8}$\\
$b\rightarrow\Omega_{cbb}$&&50.30&34.77&47.78&&52.34&$7.43\times10^{-8}$\\
$c\rightarrow\Omega_{cbb}$&&1.14&1.38&1.47&&1.49&$5.41\times10^{-10}$\\
\end{tabular}
\end{ruledtabular}
\end{table*}

The total cross sections are listed in table II for the chosen
scales. They range from a few nb to a few pb. The decimal places
are not realy significant. They are kept only for the matter of
comparison. A short look at table II reveals that although the
total cross section for some of the triply heavy baryons are small
indeed (order of pb) and their production needs energetic hadron
colliders, some others such as $b\rightarrow \Omega_{ccb}$ and
$c\rightarrow \Omega_{ccc}$ do possess larger cross sections of
the order of nb and may easily be produced at the Tevatron as
well. An interesting point in table II is that although the total
cross section for some of the particles such as $c\rightarrow
\Omega_{ccb}$ and $c\rightarrow \Omega_{cbb}$ increase with
increasing $\mu$, but this is not the case for the rest. Our
investigation shows that this depends on the range of $\mu$
selected and also on the choice of ${p_T}^{\rm cut}$ [20]. We have
also calculated the ratio $\sigma(\Omega)/\sigma(Q)$ for different
cases. The results appear in the last column of table II. Our
evaluation of charm and bottom cross sections at the LHC are
0.25497 mb and 0.46812 mb respectively.

The fragmentation production of doubly heavy baryons studied in
[4] by Doncheski {\it et al} is interesting in relation to our
work. First of all the fact that the Tevatron gives large cross
section for charm production is reflected in this work. They have
obtained nearly equal cross sections for $\Xi_{cc}$ at the
Tevatron and at the LHC. However for $\Xi_{cb}$ and $\Xi_{bb}$ the
cross sections are different. They report 430 pb, 215 pb and 16 pb
for the Tevatron and 470 pb, 490 pb and 36 pb for the LHC
respectively for these states. Although states which we have
studied are different, but physically our results are comparable
with the above.

We would like at the end discuss the uncertainties of our results.
The choice of quark masses will not only alter the fragmentation
probabilities, but also the value of $\mu$ and values of $x$ at
which the parton distribution functions are evaluated. This will
of course be reflected on the total cross sections. We have chosen
$m_c=1.25$ GeV and $m_b=4.25$ GeV which are the optimum values
reported [19]. However the slightly higher values of $m_c=1.5$ GeV
and $m_b=4.7$ GeV are also used in the literature. Changes in
quark mass will affect the fragmentation functions. In the scheme
of our calculation, the fragmentation functions inversely depend
on quark mass squared. Therefore increase in quark mass will
decrease the probabilities. The other quantity which may depend on
quark mass is the baryon decay constant. However the later is not
much clear in the case of triply heavy baryons. Taking the
explicit mass dependence of our fragmentation functions, we have
obtained 18 percent decrease in the cross sections in average,
when we use the above mentioned higher values.

There is no data on the baryon decay constant. Theoretically one
may solve the Schr\"{o}dinger like equation to obtain the wave
function at the origin for these composite particles with heavy
constituents and then relate the wave function at the origin to
the baryon decay constant. We have avoided this procedure because
of theoretical uncertainties instead have chosen $f_B=0.25$ GeV on
phenomenological grounds. The final quantity of interest is the
color factor. We have calculated this quantity using the simple
color line counting rule and have obtained $C_F=7/6$ for our
propose.

\end{document}